\documentclass[aps,prl,reprint,superscriptaddress]{revtex4-1}
\usepackage{graphicx}
\usepackage{dcolumn}
\usepackage{bm}
\usepackage{amssymb}
\usepackage{amsmath}
\usepackage{amsfonts}
\usepackage{epsf}
\usepackage{epsfig}
\usepackage{color}
\usepackage{siunitx}
\usepackage{epstopdf}

\begin{document}
\title{
\large
Separating Dipole and Quadrupole Contributions to Single-Photon Double Ionization}

\author{S.~Grundmann}
\email{grundmann@atom.uni-frankfurt.de}
\author{F.~Trinter}
\affiliation{Institut f\"ur Kernphysik, Goethe-Universit\"at, Max-von-Laue-Strasse 1, 60438 Frankfurt, Germany \\}
\author{A.~W.~Bray}
\affiliation{Research School of Physics, Australian National University, Canberra, ACT 2601, Australia \\}
\author{S.~Eckart}
\author{J.~Rist}
\author{G.~Kastirke}
\author{D.~Metz}
\affiliation{Institut f\"ur Kernphysik, Goethe-Universit\"at, Max-von-Laue-Strasse 1, 60438 Frankfurt, Germany \\}
\author{S.~Klumpp}
\affiliation{FS-FLASH-D, Deutsches Elektronen-Synchrotron (DESY), Notkestrasse 85, 22607 Hamburg, Germany \\}
\author{J.~Viefhaus}
\affiliation{Helmholtz-Zentrum Berlin, Albert-Einstein-Strasse 15, 12489 Berlin, Germany \\}
\author{L.~Ph.~H.~Schmidt}
\affiliation{Institut f\"ur Kernphysik, Goethe-Universit\"at, Max-von-Laue-Strasse 1, 60438 Frankfurt, Germany \\}
\author{J. B. Williams}
\affiliation{Department of Physics, University of Nevada, Reno, Nevada 89557, USA \\}
\author{R.~D\"orner}
\author{T.~Jahnke}
\author{M.~S.~Sch\"offler}
\affiliation{Institut f\"ur Kernphysik, Goethe-Universit\"at, Max-von-Laue-Strasse 1, 60438 Frankfurt, Germany \\}
\author{A.~S.~Kheifets}
\email{a.kheifets@anu.edu.au}
\affiliation{Research School of Physics, Australian National University, Canberra, ACT 2601, Australia \\}

\date{\today}

\begin{abstract}
We report on a kinematically complete measurement of double ionization of helium by a single 1100 eV circularly polarized photon. By exploiting dipole selection rules in the two-electron continuum state, we observed the angular emission pattern of electrons originating from a pure quadrupole transition. Our fully differential experimental data and companion ab initio nonperturbative theory show the separation of dipole and quadrupole contributions to photo-double-ionization and provide new insight into the nature of the quasifree mechanism.  
\end{abstract}

\maketitle

The interaction of photons with atoms and molecules is dominated by electronic dipole transitions due to the photon spin. Any transfer of additional orbital angular momentum arises from the photon's linear momentum $\bf{k}_\gamma$ and is consequently suppressed for low photon energies. Whenever a transition leads to the continuum, i.e., to the ejection of one or more electrons, the angular momentum becomes observable in their outgoing angular distributions. These angular distributions result from a coherent superposition of the different multipole contributions as the various angular momentum states of a free particle are energetically degenerate. In most cases however, the angular distributions are, due to the dominance of the dipole contribution, only slightly modified by the interference term between the quadrupole and the dipole transition (see \cite{hemmers} for a review). As such, the quadrupole transition amplitude alone has not been directly observed until now.

In the present work we succeeded in experimentally isolating the quadrupole contribution to photo-double-ionization [(PDI), always refers to the one-photon process in this Letter] and visualize a pure quadrupole pattern in the angular distribution of electrons emitted from a helium atom (Fig. 1). The quadrupole contribution to a photoionization process can be accessed in cases where the dominating dipole contribution is strongly suppressed \cite{Krassig2002}. For the case of double ionization, the selection rules for the two-electron continuum, which have been presented in detail by Maulbetsch and Briggs \cite{SelectionRules}, can be exploited \cite{gal, iso2004}. The most prominent of these selection rules states that for two electrons of opposite spin the electron pair wave function vanishes for total angular momentum $L$=$\hbar$ and $\bf{k}_\textit{a}=-\bf{k}_\textit{b}$ (where $\bf{k}_\text{a,b}$ are the momentum vectors of the two electrons a and b). This pattern corresponds to a nucleus at rest and two electrons receding back-to-back with equal energies. At large distances, this resembles a spatial configuration of the three charges which has only a quadrupole but no charge dipole moment. Consequently, this configuration cannot be reached by a dipole transition. Only if the energy sharing becomes unbalanced, the spatial charge configuration acquires a dipole moment. Thus, back-to-back emission with unequal energy sharing is allowed for a dipole transition.

At moderate photon energies, the strict dipole selection rule leads to a node in the electron angular distributions for helium PDI, which has been observed in the pioneering experiment by Schwarzkopf \textit{et al.} \cite{btob1} and other subsequent work (e.g. \cite{knapp_2005}). This selection rule, however, holds true only for the $L$=$\hbar$ component of the two-electron wave function. Thus, the quadrupole components can be observed directly by selecting electron pairs with opposite momentum of equal magnitude ($\bf{k}_\textit{a}=-\bf{k}_\textit{b}$) as suggested in \cite{gal}. It is by this method that we have isolated the quadrupole distribution given in Fig. 1. For the remainder of this Letter, we provide a brief outline of the experiment and the ab initio theory and then discuss in more detail how the dipole and quadrupole contributions separate in fully differential cross sections (FDCS) of which Fig. 1 is a special case.

\begin{figure} [htbp]
\centering
\includegraphics[width=1.\columnwidth]{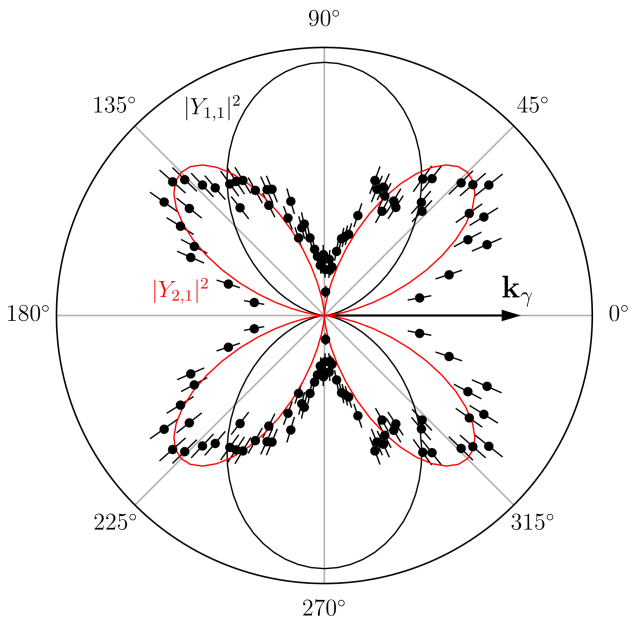}
\caption{Angular distribution of one of the electrons from photo-double-ionization of He by a single 1100 eV circularly polarized photon. The light propagation axis is horizontal ($\bf{k}_\gamma$). Data points: electrons of equal energy ($\Delta E=0.5\pm 0.1$) emitted back-to-back ($\Delta \vartheta=180 \pm 20^\circ$). For this selection, dipole contributions to the cross section vanish due to selection rules. Black line: dipole distribution ($|Y_{l=1,m=1}|^2$, not internormalized with data or quadrupole distribution), red line: quadrupole distribution fitted to the data points ($|Y_{l=2,m=1}|^2$).} 
\label{fig1}
\end{figure}

We employed a COLTRIMS reaction microscope (Cold Target Recoil Ion Momentum Spectroscopy \cite{coltrims1,coltrims2,coltrims3}) and intersected a supersonic helium gas jet with a synchrotron beam of 1100 eV circularly polarized photons from beam line P04 at PETRA III (DESY, Hamburg \cite{p04}). Electrons and ions were guided by a weak electric field (20.1 V/cm) towards two time- and position-sensitive detectors \cite{otti_1,otti_2}. Additionally, a strong magnetic field (40 G) was applied in order to confine the high-energetic electrons inside the spectrometer. An electrostatic lens and a drift tube of 80 cm length were used in the ion arm of the spectrometer to increase its momentum resolution. The high photon flux of beam line P04 combined with a dense target ($\approx$ 3$\cdot$10$^{11}$ atoms/cm$^{2}$) lead to approximately 10 millions of coincidently measured double-ionization events.

The experiment is accompanied by ab initio nonperturbative calculations using the convergent close-coupling (CCC) technique. This technique has already demonstrated its utility in identifying various mechanisms behind helium PDI at high photon energies \cite{knapp2002,kheifets_alone}. In the present work, we employed it to calculate various differential and total integrated cross sections due to dipole and quadrupole transitions. The resulting total integrated cross sections (TICS) are listed in Table 1. As a trade-off between the relative quadrupole cross section (which increases with rising photon energy) and available photon flux at beam line P04, we chose a photon energy of 1100 eV in order to perform the measurement. Circularly polarized photons were employed, because beam line P04 is currently not able to generate linearly polarized light.

\begingroup
\setlength{\tabcolsep}{10pt} 
\renewcommand{\arraystretch}{1.5} 
\begin{table*}[t]
  \centering
    \caption{Total integrated cross sections (TICS) for helium PDI as obtained from the convergent close-coupling technique compared to time-dependent close-coupling calculations from \cite{colgan}. With rising photon energy the ratio of quadrupole to dipole TICS increases. The small effect of the interference term on the cross section is neglected in this work.}
\begin{tabular}{c c c c c c c}
 \hline 
 Photon energy & \multicolumn{2}{c}{Dipole (barn)} & \multicolumn{2}{c}{Quadrupole (barn)} & \multicolumn{2}{c}{Ratio (\%)} \\ 
 • & This work & Ref. \cite{colgan} & This work & Ref. \cite{colgan} & This work & Ref. \cite{colgan}\\ 
 \hline 
 800 eV & 19.17 & 19.18 & 1.283 & 1.21 & 6.6 & 6.3 \\ 
 1100 eV & 7.011 & • & 0.6356 & • & 9.1 & • \\ 
 \hline 
\end{tabular}
  \label{tab:1}
\end{table*}  
\endgroup

In our experiment, the momenta of all the reaction products are measured in coincidence. In the case of PDI, energy and momentum conservation reduce the nine momentum components of the three particles in the final state to five independent variables. Together with cylindrical symmetry of the circularly polarized light, this makes the fully differential cross section have a fourfold dependency. By integrating over some of the remaining independent observables, we create singly (SDCS) and doubly differential cross sections (DDCS) that highlight specific features. As done so in Fig. 2, where we depict the energy sharing between the two electrons $\Delta E=\frac{E_a}{E_a+E_b}$ and the relative emission angle $\Delta \vartheta$ between them. The measured SDCS $\frac{d\sigma}{d\Delta E}$ for all electron pairs [Fig. 2(a), red dots] exhibits a very deep U-shape, indicating that the most likely energy sharing configuration consists of one electron obtaining most of the photon's energy while sharing only a small fraction with the second electron. The experimental data are normalized to the sum of dipole and quadrupole SDCS obtained from CCC calculations [Fig. 2(a), green and blue lines]. These calculations display some small numerical oscillations due to discretization of the photoelectron continuum \cite{ccc_error}. These oscillations average out in the TICS which is free from any numerical errors.  

The dominance of strong unequal energy sharing at the high photon energy of 1100 eV is a consequence of the interplay of the two established PDI mechanisms ``knock-out", also known as ``two-step-one" (TS1), and ``shake-off" (SO) \cite{knapp2002, mcguire}. In the case of a quasi-instantaneous removal of the first electron, the second electron cannot relax adiabatically to the singly charged ground state. Instead, it can be that the electron is shaken off to the continuum. For this shake-off process, small energy transfer, i.e., a very unequal energy sharing, is strongly favored. The probability for this process is determined solely by the overlap integral of the initial neutral He and final He$^+$ bound wave functions. The knock-out process is characterized by a binary collision event between the two electrons and contributes only to a small fraction of PDI events involving 1100 eV photons \cite{kheifets_alone}.

The binary collision leads to an angle of 90$^{\circ}$ between the momentum vectors of the outgoing collision partners and arbitrary energy sharing \cite{knapp2002}. In Fig. 2(b) we plot the doubly differential cross section $\frac{d^2\sigma(\Delta E, \Delta \vartheta) }{d\Delta E d\Delta \vartheta}$ as function of the relative angle between the two electrons $\Delta \vartheta$ for equal energy sharing, i.e., $\Delta E=0.5 $, where SO is strongly suppressed. The distribution is narrowly peaked at $\pm$90$^\circ$ as expected for a violent binary TS1 collision. Additionally, a distinct peak for back-to-back emission is visible, located at $\Delta \vartheta =$180$^\circ$ (the position of the node enforced by the dipole selection rule).

By restricting the measured data set to electron pairs occurring within this peak we obtain the laboratory frame electron angular distribution shown in Fig. 1. This subset of the data is, by virtue of the dipole selection rule, free of any otherwise dominating dipole contributions. Accordingly, Fig. 1 beautifully exhibits the angular distribution of a pure quadrupole transition.

As we chose the photon propagation $\bf{k}_\gamma$ to be along the quantization axis, we have the shape of the dipole distribution given by the square of the spherical harmonic $|Y_{l=1,m=1}|^2$ (black line in Fig. 1). In a quadrupole transition, the additional quantum of (orbital) angular momentum is transferred by coupling the photon's linear momentum $\bf{k}_\gamma$ to the electron. Classically, this corresponds to an angular momentum of $\bf{k}_\gamma \times \bf{r}$ which is directed perpendicularly to the light propagation ($\bf{r}$ is the electron's position vector). Hence, it increases the magnitude of the electrons angular momentum $l$ but has no effect on its projection $m$ onto $\bf{k}_\gamma$. Accordingly, the pure quadrupole contribution yields an angular distribution proportional to $|Y_{l=2,m=1}|^2$ (red line in Fig. 1). 

In terms of reaction mechanisms, the back-to-back emission at equal energy sharing is the fingerprint predicted for a route to double ionization termed ``quasifree mechanism" (QFM) \cite{amusia}, which is dipole forbidden. In the case of QFM, the nucleus is only a spectator to the photoabsorption process receiving no momentum \cite{spectator1,spectator2}. Instead, the two electrons entirely compensate each others' momentum. Our experiment confirms the existence of these ions with close to zero momentum (not shown) which have been observed by Sch\"offler \textit{et al.} in \cite{qfm_markus} at first. The probability of such events is given by the black dots in Fig. 2(a), which show the energy sharing distribution of electrons being emitted back-to-back, i.e., $\frac{d^2\sigma(\Delta E, \Delta \vartheta) }{d\Delta E d\Delta\vartheta}$ as function of  $\Delta E$ at $\Delta\vartheta =$180$^\circ \pm$12$^\circ$. As predicted by theory \cite{amusia, colgan}, this distribution has a W-shape that is similar to the quadrupole SDCS calculated using the CCC technique [Fig. 2(a), blue line]. The difference in magnitude between quadrupole SDCS and the measured DDCS in Fig. 2(a) affirms that QFM is only a fraction of the total quadrupole contribution to helium PDI.

\begin{figure}[t]
\centering
\includegraphics[width=1.\columnwidth]{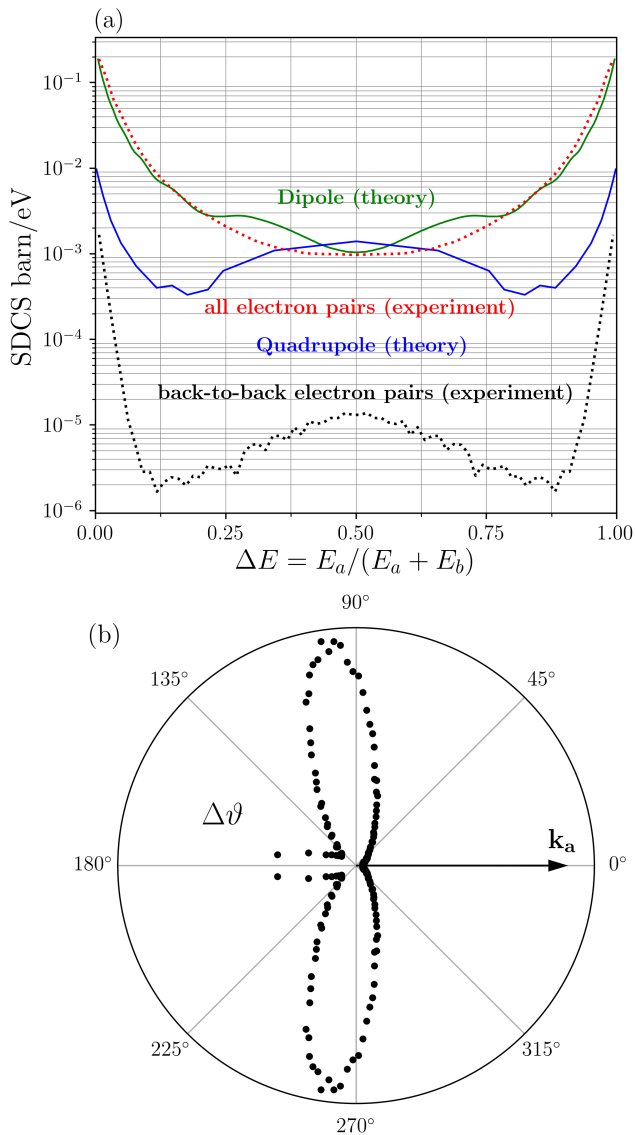}
\caption{(a) Red dots: singly differential cross section (SDCS) for photo-double-ionization of helium by a single 1100 eV circularly polarized photon as function of the electron energy sharing $\Delta E$. Black dots: doubly differential cross section $\frac{d^2\sigma(\Delta E, \Delta \vartheta) }{d\Delta E d\Delta\vartheta}$ for electrons emitted back-to-back ($\Delta \vartheta=180^\circ \pm 12^\circ$) as a function of $\Delta E $. Green (blue) line: dipole (quadrupole) SDCS from CCC calculations. The red dots are normalized to the sum of the dipole and quadrupole SDCSs. (b) Distribution of the relative emission angle $\Delta \vartheta$ between both electrons for equal energy sharing, i.e., doubly differential  cross section $\frac{d^2\sigma(\Delta E, \Delta \vartheta) }{d\Delta E d\Delta\vartheta}$ for $\Delta E=0.5 \pm 0.1 $ as function of $\Delta \vartheta $. Error bars are smaller than dot size in (a) and (b).} 
\label{fig2}
\end{figure}

The most complete picture of the double ionization process is provided by fully differential cross sections (FDCS) $\frac{d^4 \sigma(\vartheta_a, \vartheta_b,\Phi_{ab},\Delta E)}{d\vartheta_a d\vartheta_b d\Phi_{ab} d\Delta E }$. Here $\vartheta_{a,b}$ denote the polar angles of the two electrons with respect to $\bf{k}_\gamma$ and $\Phi_{ab}$ labels the difference between their respective azimuthal angles, i.e., the angles around the light propagation axis. We inspect the coplanar geometry where $\bf{k}_\gamma$, $\bf{k}_\textit{a}$, and $\bf{k}_\textit{b}$ are all in one plane as $\Phi_{ab}=0^\circ$, $180^\circ$.

A suitable parametrization of the transition amplitude of helium PDI with electrons confined to this coplanar geometry has been presented in \cite{iso2005_71}. This parametrization separates the angular dependence of the transition amplitude from the energy dependence and the dipole component $A_d$ from the quadrupole component $A_q$. While the dipole contribution has the form
\begin{equation*}
A_d=f_a \text{sin} \vartheta_a + f_b \text{sin} \vartheta_b \text{ ,}
\end{equation*}
the quadrupole fraction of the amplitude reads as
\begin{align*}
A_q&=g_a \text{cos} \vartheta_a \text{sin} \vartheta_a + g_b \text{cos} \vartheta_b \text{sin} \vartheta_b \\
&+ g_s ( \text{cos} \vartheta_a \text{sin} \vartheta_b+ \text{cos} \vartheta_b \text{sin} \vartheta_a ) \text{ .} 
\end{align*}
The dynamic factors $f_a$, $f_b$, $g_a$, $g_b$, and $g_s$ depend on the electron mutual angle $\Delta \vartheta$ and the electron energy sharing $\Delta E$. While they are described comprehensively in \cite{iso2005_71}, it is noteworthy that the parallel emission of the two electrons, i.e., $\Delta \vartheta=0$, is strongly suppressed by these factors. At equal energy sharing, $f_a$ and $f_b$ are identical and
\begin{equation*}
A_d \propto \text{sin} \vartheta_a + \text{sin} \vartheta_b = \text{sin} \left( \frac{\vartheta_a + \vartheta_b}{2} \right) \text{cos} \left( \frac{\vartheta_a - \vartheta_b}{2} \right) \text{ .} 
\end{equation*}
Consequently, the dipole amplitude in the coplanar geometry vanishes, if
\begin{equation*}
|\vartheta_a - \vartheta_b|=(2n+1)\pi \text{ } \lor \text{ } \vartheta_a + \vartheta_b = 2n \pi \text{ .}
\end{equation*}
In the case of back-to-back emission, the first condition is always satisfied. Consequently, this analysis of the angular factors alone demonstrates how the back-to-back emission with equal energy sharing is dipole forbidden. Unlike the dipole amplitude however, the quadrupole component allows the back-to-back emission at equal energy sharing as
\begin{equation*}
A_q \propto g_m \text{cos} \vartheta_a \text{sin} \vartheta_a \propto Y_{l=2,m=1} \text{ ,}
\end{equation*}
with $g_a=g_b$ and $g_m=2g_a+2g_s$. Thus, the squared quadrupole amplitude $|A_q|^2$ possesses the characteristic fourfold symmetry clearly visible in Fig. 1.

Figure 3 presents the fully differential cross section restricted to the coplanar geometry and to equal energy sharing. Figure 3(a) shows the dipole contributions to the FDCS as obtained from CCC calculations while Fig. 3(b) contains the results of such calculations for the quadrupole term. We see that due to entirely different symmetries, dipole and quadrupole contributions to the FDCS are completely separated. Emphasis should be put on the fact that in this result the lowest order contribution of the quadrupole term comes from $|A_q|^2$ but not from the interference term $|A_d A_q|$ as, for example, in \cite{spectator2}. These CCC predictions are in excellent agreement with the experimental results in Fig. 3(c). The measured distribution can clearly be identified as the superposition of Figs. 3(a) and 3(b). 

\begin{figure*} [t]
\centering
\includegraphics[width=2.\columnwidth]{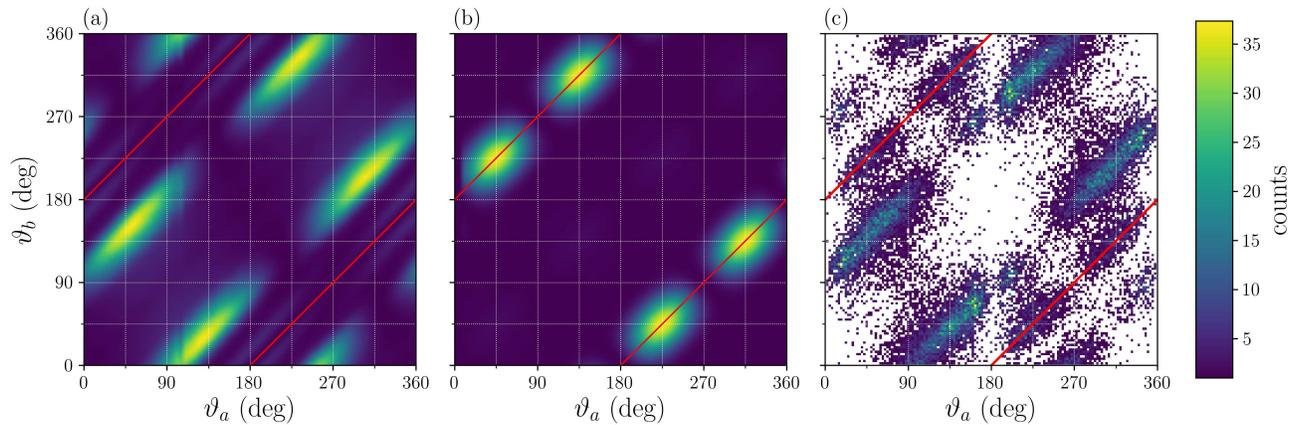}
\caption{Fully differential cross section for photo-double-ionization of He by a single 1100 eV circularly polarized photon, $\frac{d^4 \sigma(\vartheta_a, \vartheta_b,\Phi_{ab},\Delta E)}{d\vartheta_a d\vartheta_b d\Phi_{ab} d\Delta E }$, for coplanar geometry ($\bf{k}_{a,b}$ and $\bf{k}_\gamma$ in one plane) and equal energy sharing. The solid red lines visualize the conditions under which the dipole amplitude vanishes in a coplanar geometry. Dipole (a) and quadrupole (b) contributions to the FDCS as obtained from CCC calculations (not internormalized). The measured distribution (c) shows separated dipole and quadrupole contributions with the gates $\Phi_{ab}=0^\circ\pm 20^\circ,180^\circ \pm 20^\circ$, and $\Delta E =0.5 \pm 0.1$ and it can be identified as the superposition of (a) and (b). $\vartheta_{a,b}=0^\circ$ correspond to emission in the direction of photon propagation.}
\label{fig3}
\end{figure*}

In conclusion, we have unambiguously separated the quadrupole contribution to photo-double-ionization. We find a clean quadrupolar angular distribution in the laboratory frame for electrons that have been emitted back-to-back with equal energy. This is the first work of its kind where quadrupole effects have been shown with such demonstrable clarity. Furthermore, we have finally and unambiguously identified all predicted fingerprints of the QFM in various observables of helium PDI at a high photon energy. Our measured fully differential cross sections are in excellent agreement with the calculations of the CCC theory. In the future, we plan to conduct similar measurements on various targets by exploiting dipole-forbidden but quadrupole-allowed kinematics. In complex molecules, non-dipole photoionization may become a sensitive probe of electron localization.

\acknowledgments   A. K. acknowledges support by the Wilhelm and Else Heraeus Foundation. S. K. acknowledges the funding of the EUCALL project within the \textit{European Union’s Horizon 2020 research and innovation programme} under the grant agreement No. 654220. This work was supported by BMBF and DFG. We are grateful to the staff of PETRA III for excellent support during the beam
time.

\bibliographystyle{apsrev4-1}

\begin{thebibliography}{16}%
	\makeatletter
	\providecommand \@ifxundefined [1]{%
		\@ifx{#1\undefined}
	}%
	\providecommand \@ifnum [1]{%
		\ifnum #1\expandafter \@firstoftwo
		\else \expandafter \@secondoftwo
		\fi
	}%
	\providecommand \@ifx [1]{%
		\ifx #1\expandafter \@firstoftwo
		\else \expandafter \@secondoftwo
		\fi
	}%
	\providecommand \natexlab [1]{#1}%
	\providecommand \enquote  [1]{``#1''}%
	\providecommand \bibnamefont  [1]{#1}%
	\providecommand \bibfnamefont [1]{#1}%
	\providecommand \citenamefont [1]{#1}%
	\providecommand \href@noop [0]{\@secondoftwo}%
	\providecommand \href [0]{\begingroup \@sanitize@url \@href}%
	\providecommand \@href[1]{\@@startlink{#1}\@@href}%
	\providecommand \@@href[1]{\endgroup#1\@@endlink}%
	\providecommand \@sanitize@url [0]{\catcode `\\12\catcode `\$12\catcode
		`\&12\catcode `\#12\catcode `\^12\catcode `\_12\catcode `\%12\relax}%
	\providecommand \@@startlink[1]{}%
	\providecommand \@@endlink[0]{}%
	\providecommand \url  [0]{\begingroup\@sanitize@url \@url }%
	\providecommand \@url [1]{\endgroup\@href {#1}{\urlprefix }}%
	\providecommand \urlprefix  [0]{URL }%
	\providecommand \Eprint [0]{\href }%
	\providecommand \doibase [0]{http://dx.doi.org/}%
	\providecommand \selectlanguage [0]{\@gobble}%
	\providecommand \bibinfo  [0]{\@secondoftwo}%
	\providecommand \bibfield  [0]{\@secondoftwo}%
	\providecommand \translation [1]{[#1]}%
	\providecommand \BibitemOpen [0]{}%
	\providecommand \bibitemStop [0]{}%
	\providecommand \bibitemNoStop [0]{.\EOS\space}%
	\providecommand \EOS [0]{\spacefactor3000\relax}%
	\providecommand \BibitemShut  [1]{\csname bibitem#1\endcsname}%
	\let\auto@bib@innerbib\@empty
\bibitem {hemmers}%
	\BibitemOpen
	\bibfield  {author} {
	\bibinfo {author} {\bibfnamefont {O. Hemmers}},
	\bibinfo {author} {\bibfnamefont {R. Guillemin}},
	\ and\ \bibinfo {author} \bibnamefont {D. W. Lindle},\ }\href {https://doi.org/10.1016/j.radphyschem.2003.12.009}{\bibfield {journal}{	\bibinfo {journal} {Rad. Phys. Chem.}\ }
	\textbf {\bibinfo {volume} {70}},\
	\bibinfo {pages} {123}
	(\bibinfo {year} {2004})}
	\BibitemShut {NoStop}%
	\bibitem {Krassig2002}%
	\BibitemOpen
	\bibfield  {author} {
	\bibinfo {author} {\bibfnamefont {B. Kr\"assig}},
	\bibinfo {author} {\bibfnamefont {E. P. Kanter}},
	\bibinfo {author} {\bibfnamefont {S. H. Southworth}},
	\bibinfo {author} {\bibfnamefont {R. Guillemin}},
	\bibinfo {author} {\bibfnamefont {O. Hemmers}},
	\bibinfo {author} {\bibfnamefont {D. W. Lindle}},
	\bibinfo {author} {\bibfnamefont {R. Wehlitz}},
	\ and\ \bibinfo {author} \bibnamefont {N. L. S. Martin},\ }\href {https://link.aps.org/doi/10.1103/PhysRevLett.88.203002}{\bibfield {journal}{	\bibinfo {journal} {Phys. Rev. Lett.}\ }
	\textbf {\bibinfo {volume} {88}},\
	\bibinfo {pages} {203002}
	(\bibinfo {year} {2002})}
	\BibitemShut {NoStop}%
	\bibitem {SelectionRules}%
	\BibitemOpen
	\bibfield  {author} {
	\bibinfo {author} {\bibfnamefont {F. Maulbetsch}},
	\ and\ \bibinfo {author} \bibnamefont {J. S. Briggs},\ }\href {http://stacks.iop.org/0953-4075/28/i=4/a=007}{\bibfield {journal}{	\bibinfo {journal} {J. Phys. B}\ }
	\textbf {\bibinfo {volume} {28}},\
	\bibinfo {pages} {551}
	(\bibinfo {year} {1995})}
	\BibitemShut {NoStop}%
\bibitem {gal}%
	\BibitemOpen
	\bibfield  {author} {
	\bibinfo {author} {\bibfnamefont {A. G. Galstyan}},
	\bibinfo {author} {\bibfnamefont {O. Chuluunbaatar}},
	\bibinfo {author} {\bibfnamefont {Y. V. Popov}},
	\ and\ \bibinfo {author} \bibnamefont {B. Piraux},\ }\href {https://doi.org/10.1103/PhysRevA.85.023418}{\bibfield {journal}{	\bibinfo {journal} {Phys. Rev. A}\ }
	\textbf {\bibinfo {volume} {85}},\
	\bibinfo {pages} {023418}
	(\bibinfo {year} {2012})}
	\BibitemShut {NoStop}%
	\bibitem {iso2004}%
	\BibitemOpen
	\bibfield  {author} {
	\bibinfo {author} {\bibfnamefont {A. Y. Istomin}},
	\bibinfo {author} {\bibfnamefont {N. L. Manakov}},
	\bibinfo {author} {\bibfnamefont {A. V. Meremianin}},
	\ and\ \bibinfo {author} \bibnamefont {A. F. Starace},\ }\href {https://doi.org/10.1103/PhysRevLett.92.063002}{\bibfield {journal}{	\bibinfo {journal} {Phys. Rev. Lett.}\ }
	\textbf {\bibinfo {volume} {92}},\
	\bibinfo {pages} {063002}
	(\bibinfo {year} {2004})}
	\BibitemShut {NoStop}%
\bibitem {btob1}%
	\BibitemOpen
	\bibfield  {author} {
	\bibinfo {author} {\bibfnamefont {O. Schwarzkopf}},
	\bibinfo {author} {\bibfnamefont {B. Kr\"assig}},
	\bibinfo {author} {\bibfnamefont {J. Elmiger}},
	\ and\ \bibinfo {author} \bibnamefont {V. Schmidt},\ }\href {https://doi.org/10.1103/PhysRevLett.70.3008}{\bibfield {journal}{	\bibinfo {journal} {Phys. Rev. Lett.}\ }
	\textbf {\bibinfo {volume} {70}},\
	\bibinfo {pages} {3008}
	(\bibinfo {year} {1993})}
	\BibitemShut {NoStop}%
\bibitem {knapp_2005}%
	\BibitemOpen
	\bibfield  {author} {
	\bibinfo {author} {\bibfnamefont {A. Knapp}},
	\bibinfo {author} {\bibfnamefont {A. Kheifets}},
	\bibinfo {author} {\bibfnamefont {I. Bray}},
	\bibinfo {author} {\bibfnamefont {Th. Weber}},
	\bibinfo {author} {\bibfnamefont {A. L. Landers}},
	\bibinfo {author} {\bibfnamefont {S. Sch\"ossler}},
	\bibinfo {author} {\bibfnamefont {T. Jahnke}},
	\bibinfo {author} {\bibfnamefont {J. Nickles}},
	\bibinfo {author} {\bibfnamefont {S. Kammer}},
	\bibinfo {author} {\bibfnamefont {O. Jagutzki}},
	\bibinfo {author} {\bibfnamefont {L. Ph. H. Schmidt}},
	\bibinfo {author} {\bibfnamefont {M. Sch\"offler}},
	\bibinfo {author} {\bibfnamefont {T. Osipov}},
	\bibinfo {author} {\bibfnamefont {M. H. Prior}},
	\bibinfo {author} {\bibfnamefont {H. Schmidt-B\"ocking}},
	\bibinfo {author} {\bibfnamefont {C. L. Cocke}},
	\ and\ \bibinfo {author} \bibnamefont {R. D\"orner},\ }\href {https://doi.org/10.1088/0953-4075/38/6/001} {\bibfield {journal}{	\bibinfo {journal} {J. Phys. B}\ }
	\textbf {\bibinfo {volume} {38}},\
	\bibinfo {pages} {615}
	(\bibinfo {year} {2005})}
	\BibitemShut {NoStop}%
\bibitem {coltrims1}%
	\BibitemOpen
	\bibfield  {author} {
	\bibinfo {author} {\bibfnamefont {R. D\"orner}},
	\bibinfo {author} {\bibfnamefont {V. Mergel}},
	\bibinfo {author} {\bibfnamefont {O. Jagutzki}},
	\bibinfo {author} {\bibfnamefont {L. Spielberger}},
	\bibinfo {author} {\bibfnamefont {J. Ullrich}},
	\bibinfo {author} {\bibfnamefont {R. Moshammer}},
	\ and\ \bibinfo {author} \bibnamefont {H. Schmidt-B\"ocking},\ }\href {https://doi.org/10.1016/S0370-1573(99)00109-X}{\bibfield {journal}{	\bibinfo {journal} {Phys. Rep.}\ }
	\textbf {\bibinfo {volume} {330}},\
	\bibinfo {pages} {95}
	(\bibinfo {year} {2000})}
	\BibitemShut {NoStop}%
\bibitem {coltrims2}%
	\BibitemOpen
	\bibfield  {author} {
	\bibinfo {author} {\bibfnamefont {J. Ullrich}},
	\bibinfo {author} {\bibfnamefont {R. Moshammer}},
	\bibinfo {author} {\bibfnamefont {A. Dorn}},	
	\bibinfo {author} {\bibfnamefont {R. D\"orner}},
	\bibinfo {author} {\bibfnamefont {L. Ph. H. Schmidt}},
	\ and\ \bibinfo {author} \bibnamefont {H. Schmidt-B\"ocking},\ }\href {http://stacks.iop.org/0034-4885/66/i=9/a=203}{\bibfield {journal}{	\bibinfo {journal} {Rep. Prog. Phys.}\ }
	\textbf {\bibinfo {volume} {66}},\
	\bibinfo {pages} {1463}
	(\bibinfo {year} {2003})}
	\BibitemShut {NoStop}%
\bibitem {coltrims3}%
	\BibitemOpen
	\bibfield  {author} {
	\bibinfo {author} {\bibfnamefont {T. Jahnke}},
	\bibinfo {author} {\bibfnamefont {Th. Weber}},
	\bibinfo {author} {\bibfnamefont {T. Osipov}},
	\bibinfo {author} {\bibfnamefont {O. Jagutzki}}
	\bibinfo {author} {\bibfnamefont {L. Ph. H. Schmidt}},
	\bibinfo {author} {\bibfnamefont {C. L. Cocke}},
	\bibinfo {author} {\bibfnamefont {M. H. Prior}},
	\bibinfo {author} {\bibfnamefont {H. Schmidt-B\"ocking}},
	\ and\ \bibinfo {author} \bibnamefont {R. D\"orner},\ }\href {https://doi.org/10.1016/j.elspec.2004.06.010}{\bibfield {journal}{	\bibinfo {journal} {J. Electron Spectrosc. Relat. Phenom.}\ }
	\textbf {\bibinfo {volume} {141}},\
	\bibinfo {pages} {229}
	(\bibinfo {year} {2004})}
	\BibitemShut {NoStop}%
\bibitem {p04}%
	\BibitemOpen
	\bibfield  {author} {
	\bibinfo {author} {\bibfnamefont {J. Viefhaus}},
	\bibinfo {author} {\bibfnamefont {F. Scholz}},
	\bibinfo {author} {\bibfnamefont {S. Deinert}},
	\bibinfo {author} {\bibfnamefont {L. Glaser}}
	\bibinfo {author} {\bibfnamefont {M. Ilchen}},
	\bibinfo {author} {\bibfnamefont {J. Seltmann}},
	\bibinfo {author} {\bibfnamefont {P. Walter}},
	\ and\ \bibinfo {author} \bibnamefont {F. Siewert},\ }\href {https://doi.org/10.1016/j.nima.2012.10.110}{\bibfield {journal}{	\bibinfo {journal} {Nucl. Instrum. Methods Phys. Res., Sect. A}\ }
	\textbf {\bibinfo {volume} {710}},\
	\bibinfo {pages} {151}
	(\bibinfo {year} {2013})}
	\BibitemShut {NoStop}%
\bibitem {otti_1}%
	\BibitemOpen
	\bibfield  {author} {
	\bibinfo {author} {\bibfnamefont {O. Jagutzki}},
	\bibinfo {author} {\bibfnamefont {J. Lapington}},
	\bibinfo {author} {\bibfnamefont {L. Worth}},
	\bibinfo {author} {\bibfnamefont {U. Spillman}}
	\bibinfo {author} {\bibfnamefont {V. Mergel}},
	\ and\ \bibinfo {author} \bibnamefont {H. Schmidt-B\"ocking},\ }\href {https://doi.org/10.1016/S0168-9002(01)01843-5}{\bibfield {journal}{	\bibinfo {journal} {Nucl. Instrum. Methods Phys. Res., Sect. A}\ }
	\textbf {\bibinfo {volume} {477}},\
	\bibinfo {pages} {256}
	(\bibinfo {year} {2002})}
	\BibitemShut {NoStop}%
\bibitem {otti_2}%
	\BibitemOpen
	\bibfield  {author} {
	\bibinfo {author} {\bibfnamefont {O. Jagutzki}},
	\bibinfo {author} {\bibfnamefont {V. Mergel}},
	\bibinfo {author} {\bibfnamefont {K. Ullmann-Pfleger}},
	\bibinfo {author} {\bibfnamefont {L. Spielberger}},
	\bibinfo {author} {\bibfnamefont {U. Spillman}}
	\bibinfo {author} {\bibfnamefont {R. D\"orner}},
	\ and\ \bibinfo {author} \bibnamefont {H. Schmidt-B\"ocking},\ }\href {https://doi.org/10.1016/S0168-9002(01)01839-3}{\bibfield {journal}{	\bibinfo {journal} {Nucl. Instrum. Methods Phys. Res., Sect. A}\ }
	\textbf {\bibinfo {volume} {477}},\
	\bibinfo {pages} {244}
	(\bibinfo {year} {2002})}
	\BibitemShut {NoStop}%
\bibitem {knapp2002}%
	\BibitemOpen
	\bibfield  {author} {
	\bibinfo {author} {\bibfnamefont {A. Knapp}},
	\bibinfo {author} {\bibfnamefont {A. S. Kheifets}},
	\bibinfo {author} {\bibfnamefont {I. Bray}},
	\bibinfo {author} {\bibfnamefont {Th. Weber}},
	\bibinfo {author} {\bibfnamefont {A. L. Landers}},
	\bibinfo {author} {\bibfnamefont {S. Sch\"ossler}},
	\bibinfo {author} {\bibfnamefont {T. Jahnke}},
	\bibinfo {author} {\bibfnamefont {J. Nickles}},
	\bibinfo {author} {\bibfnamefont {S. Kammer}},
	\bibinfo {author} {\bibfnamefont {O. Jagutzki}},
	\bibinfo {author} {\bibfnamefont {L. Ph. H. Schmidt}},
	\bibinfo {author} {\bibfnamefont {T. Osipov}},
	\bibinfo {author} {\bibfnamefont {J. R\"osch}},
	\bibinfo {author} {\bibfnamefont {M. H. Prior}},
	\bibinfo {author} {\bibfnamefont {H. Schmidt-B\"ocking}},
	\bibinfo {author} {\bibfnamefont {C. L. Cocke}},
	\ and\ \bibinfo {author} \bibnamefont {R. D\"orner},\ }\href {https://doi.org/10.1103/PhysRevLett.89.033004}{\bibfield {journal}{	\bibinfo {journal} {Phys. Rev. Lett.}\ }
	\textbf {\bibinfo {volume} {89}},\
	\bibinfo {pages} {033004}
	(\bibinfo {year} {2002})}
	\BibitemShut {NoStop}%
\bibitem {kheifets_alone}%
	\BibitemOpen
	\bibfield  {author} {
	\bibinfo {author} {\bibfnamefont {A. Kheifets}}},\href {http://iopscience.iop.org/article/10.1088/0953-4075/34/8/102/meta}{\bibfield {journal}{	\bibinfo {journal} {J. Phys. B}\ }
	\textbf {\bibinfo {volume} {34}},\
	\bibinfo {pages} {L247}
	(\bibinfo {year} {2001})}
	\BibitemShut {NoStop}%
\bibitem {ccc_error}%
	\BibitemOpen
	\bibfield  {author} {
	\bibinfo {author} {\bibfnamefont {I. Bray}},
	\bibinfo {author} {\bibfnamefont {D. V. Fursa}},
	\bibinfo {author} {\bibfnamefont {A. S. Kadyrov}},
	\bibinfo {author} {\bibfnamefont {A. T. Stelbovics}},
	\bibinfo {author} {\bibfnamefont {A. S. Kheifets}},
	\ and\ \bibinfo {author} \bibnamefont {A. M. Mukhamedzhanov},\ }\href {https://www.sciencedirect.com/science/article/pii/S0370157312002062}{\bibfield {journal}{	\bibinfo {journal} {Phys. Rep.}\ }
	\textbf {\bibinfo {volume} {520}},\
	\bibinfo {pages} {135}
	(\bibinfo {year} {2012})}
	\BibitemShut {NoStop}%
\bibitem {mcguire}%
	\BibitemOpen
	\bibfield  {author} {
	\bibinfo {author} {\bibfnamefont {J. H. McGuire}}},
	{\bibfield {journal}{	\bibinfo {journal} {Electron Correlation and Dynamics in Atomic Collisions, Cambridge University Press}\ }
	(\bibinfo {year} {2005})}
	\BibitemShut {NoStop}%
\bibitem {amusia}%
	\BibitemOpen
	\bibfield  {author} {
	\bibinfo {author} {\bibfnamefont {M. Ya. Amusia}},
	\bibinfo {author} {\bibfnamefont {E. G. Drukarev}},
	\bibinfo {author} {\bibfnamefont {V. G. Gorshkov}},
	\ and\ \bibinfo {author} \bibnamefont {M. O. Kazachkov},\ }\href {http://iopscience.iop.org/article/10.1088/0022-3700/8/8/016/pdf}{\bibfield {journal}{	\bibinfo {journal} {J. Phys. B}\ }
	\textbf {\bibinfo {volume} {8}},\
	\bibinfo {pages} {1248}
	(\bibinfo {year} {1975})}
	\BibitemShut {NoStop}%
\bibitem {spectator1}%
	\BibitemOpen
	\bibfield  {author} {
	\bibinfo {author} {\bibfnamefont {T. Suric}},
	\bibinfo {author} {\bibfnamefont {E. G. Drukarev}},
	\ and\ \bibinfo {author} \bibnamefont {R. H. Pratt},\ }\href {https://doi.org/10.1103/PhysRevA.67.022709}{\bibfield {journal}{	\bibinfo {journal} {Phys. Rev. A.}\ }
	\textbf {\bibinfo {volume} {67}},\
	\bibinfo {pages} {022709}
	(\bibinfo {year} {2003})}
	\BibitemShut {NoStop}%
\bibitem {spectator2}%
	\BibitemOpen
	\bibfield  {author} {
	\bibinfo {author} {\bibfnamefont {M. Y. Amusia}},
	\ and\ \bibinfo {author} \bibnamefont {E. G. Drukarev},\ }\href {https://doi.org/10.1088/0953-4075/36/12/304}{\bibfield {journal}{	\bibinfo {journal} {J. Phys. B}\ }
	\textbf {\bibinfo {volume} {36}},\
	\bibinfo {pages} {2433}
	(\bibinfo {year} {2003})}
	\BibitemShut {NoStop}%
\bibitem {qfm_markus}%
	\BibitemOpen
	\bibfield  {author} {
	\bibinfo {author} {\bibfnamefont {M. S. Sch\"offler}},
	\bibinfo {author} {\bibfnamefont {C. Stuck}},
	\bibinfo {author} {\bibfnamefont {M. Waitz}},
	\bibinfo {author} {\bibfnamefont {F. Trinter}},
	\bibinfo {author} {\bibfnamefont {T. Jahnke}},
	\bibinfo {author} {\bibfnamefont {U. Lenz}},
	\bibinfo {author} {\bibfnamefont {M. Jones}},
	\bibinfo {author} {\bibfnamefont {A. Belkacem}},
	\bibinfo {author} {\bibfnamefont {A. L. Landers}},
	\bibinfo {author} {\bibfnamefont {M. S. Pindzola}},
	\bibinfo {author} {\bibfnamefont {C. L. Cocke}},
	\bibinfo {author} {\bibfnamefont {J. Colgan}},
	\bibinfo {author} {\bibfnamefont {A. Kheifets}},
	\bibinfo {author} {\bibfnamefont {I. Bray}},
	\bibinfo {author} {\bibfnamefont {H. Schmidt-B\"ocking}},
	\bibinfo {author} {\bibfnamefont {R. D\"orner}},
	\ and\ \bibinfo {author} \bibnamefont {Th. Weber},\ }\href {http://link.aps.org/doi/10.1103/PhysRevLett.111.013003}{\bibfield {journal}{	\bibinfo {journal} {Phys. Rev. Lett.}\ }
	\textbf {\bibinfo {volume} {111}},\
	\bibinfo {pages} {013003}
	(\bibinfo {year} {2013})}
	\BibitemShut {NoStop}%
\bibitem {colgan}%
	\BibitemOpen
	\bibfield  {author} {
	\bibinfo {author} {\bibfnamefont {J. A. Ludlow}},
	\bibinfo {author} {\bibfnamefont {J. Colgan}},
	\bibinfo {author} {\bibfnamefont {Teck-Ghee Lee}},
	\bibinfo {author} {\bibfnamefont {M. S. Pindzola}},
	\ and\ \bibinfo {author} \bibnamefont {F. Robicheaux},\ }\href {https://doi.org/10.1088/0953-4075/42/22/225204}{\bibfield {journal}{	\bibinfo {journal} {J. Phys. B}\ }
	\textbf {\bibinfo {volume} {42}},\
	\bibinfo {pages} {225204}
	(\bibinfo {year} {2009})}
	\BibitemShut {NoStop}%
\bibitem {iso2005_71}%
	\BibitemOpen
	\bibfield  {author} {
	\bibinfo {author} {\bibfnamefont {A. Y. Istomin}},
	\bibinfo {author} {\bibfnamefont {N. L. Manakov}},
	\bibinfo {author} {\bibfnamefont {A. V. Meremianin}},
	\ and\ \bibinfo {author} \bibnamefont {A. F. Starace},\ }\href {https://doi.org/10.1103/PhysRevA.71.052702}{\bibfield {journal}{	\bibinfo {journal} {Phys. Rev. A}\ }
	\textbf {\bibinfo {volume} {71}},\
	\bibinfo {pages} {052702}
	(\bibinfo {year} {2005})}
	\BibitemShut {NoStop}%
\end{thebibliography}

\end{document}